\documentclass[aps,prl,twocolumn]{revtex4}

\usepackage{epsf}
\usepackage{array}
\newcommand{\beq}{\begin{equation}}
\newcommand{\eeq}{\end{equation}}
\newcommand{\be}{\begin{equation}}
\newcommand{\ee}{\end{equation}}
\newcommand{\beqa}{\begin{eqnarray}}
\newcommand{\eeqa}{\end{eqnarray}}
\newcommand{\bea}{\begin{eqnarray}}
\newcommand{\eea}{\end{eqnarray}}

\begin{document}
\title{On the ``non-perturbative analysis'' 
of zero-temperature dephasing:\\
I. Dyson equation and self energy.}
\author{I.L. Aleiner} 
\affiliation{Physics Department, Columbia University,
New York, NY 10027 }
\author{B.L. Altshuler} 
\affiliation{Physics Department, Princeton University,
Princeton, NJ 08545}
\author{M.G. Vavilov} 
\affiliation{Theoretical Physics Institute, University of Minnesota,
Minneapolis, MN 55455}

\date{\today}

\begin{abstract}                
We point out that the structure of the self-energy suggested in
cond-mat/0208140 as a result of a ``non-perturbative analysis'' by
``purely mathematical means'' is incompatible with the very definition
of the self-energy.
\end{abstract} 
\maketitle 

Recent paper~\cite{GZnew} by Golubev and Zaikin, called a ``reply'' to
our comment \cite{AAV}, contains nothing new but attempts to justify
and improve the calculation of the preexponential factors for some
``path integrals'' having nothing to do with the contributions
identified in the second part of Ref.~\cite{AAV}. For this reason, we
do not accept it as a  response to our objections, 
and will not analyze the  calculation of  
Ref.~\cite{GZnew} here. However, since GZ
misrepresent   basic notions of field theory
methods \cite{Dyson} rather than only their applications to disordered systems,
we find it necessary to make several comments. Below, we will show
that Eqs.~(5) and (8) of Ref.~\cite{GZnew} 
[copied below as Eqs.~(\ref{GZmain})
and (\ref{voblia})], used by GZ to explain
 ``why perturbative in the interaction techniques are insufficient
for the problem in question'', contradict the construction of 
Dyson equation.
For pedagogical reasons we recall ABC of Dyson self-energy
first, and then analyse GZ arguments.

{\it ABC of Dyson self-energy} --- According to Dyson \cite{Dyson},
any propagator, such as an electron  Green function
$G(\omega)$, or a Cooperon in the present case, is connected to its
bare form, $G_0(\omega)$, and a self energy $\Sigma(\omega)$
\begin{equation}
G(\omega)=\frac{1}{G_0^{-1}(\omega)-\Sigma(\omega)},
\label{Dyson}
\end{equation}
where $\omega$ is a schematic notation for energies and momenta of the
particle or Cooperon.  It should be emphasized
that 
$\Sigma$ is not an ``ambiguous'' quantity defined by an expression
$\Sigma=1/G_0-1/G$ but rather is a well defined mathematical object -- sum
of all {\it one-particle irreducible} graphs, see Fig.~1 (a-c) for the
lowest order contributions.  

The significance of the introduction of
$\Sigma$  is the following: the expansion
of Eq.~(\ref{Dyson}) in powers of the interaction strength contains
poles of higher and higher orders, $G^n_0(\omega)$.

In contrast, $\Sigma(\omega)$ includes one particle irreducible graphs
only and does {\em not} contain contributions proportional to
$G^n_0(\omega)$ -- each of the contributions is a non-factorizable
integral over internal momenta and energies.  These integrations
either eliminate or substantially reduce the singularities of
$\Sigma$. In particular, when the integrals are ultraviolet
(determined by large energies and momenta and proportional to some
power of the high-energy cut-off) $\Sigma(\omega)$ is finite at
$\omega\to 0$, and its expansion in powers of the interaction coupling
constant is a well defined asymptotic series.

The conventional scheme does not assume that one determines
$G(\omega)$ by some means (non-perturbative analysis) and then
evaluates $\Sigma(\omega)$ through Eq.~(\ref{Dyson}). Quite contrary, if
$G(\omega)$ can be expanded in terms of the interaction strength, each
of these terms should be also possible to obtain from the diagrammatic
expansion for $\Sigma(\omega)$. As long as the two approaches give
different results, it is the "non-perturbative analysis" rather than
the diagrammatic expansion to be questioned.

{\it Golubev-Zaikin's self-energy} -- Let us discuss the self energy of the
Cooperon, presented by GZ in Ref.~\cite{GZnew}: 
\\ (Eq.~(5) of Ref.~\cite{GZnew})
\be
\Sigma(\omega)= \frac{(\alpha+\beta T)^2-i\omega \beta T}{2\alpha+\beta T -i\omega} ,
\label{GZmain}
\ee
``where $\alpha$ and $\beta$ are proportional to interaction
strength\cite{GZnew}''.  Since $\Sigma (\omega)$ from
Eq.~(\ref{GZmain}) at any finite $\omega$ can be expanded in terms of
$\alpha$ and $\beta$, \emph{it is perturbative and should be
accessible by usual diagrammatic approach}.  On the other hand this
expansion is singular at $\omega \to 0$: 
\\ (Eq.~(8) of Ref.~\cite{GZnew})
\be
\Sigma(\omega)= \beta T + \frac{\alpha^2}{-i\omega}+\dots ,
\label{voblia}
\ee
note that each term in this expansion is proportional to a power of
interaction constant.
Once again, Eqs.~(\ref{GZmain}) and (\ref{voblia}) are the
equations used by GZ to demonstrate that the
 `` perturbative in the interaction techniques are insufficient
for the problem in question''.

In GZ scheme, $\alpha$ is determined by ultraviolet integral and
proportional to the high energy scale $1/\tau_e$, which is supposed to
be much bigger than $T, 1/\alpha$ and any other scale in the problem
\cite{footnote}. The form of
the second term in the right hand side of Eq.~(\ref{voblia}) implies
that this term \\
{\it (i)} is the second order expansion in $\alpha$ of the
self-energy, see Fig. 1(b-c);
\\
{\it (ii)} contains the same pole as bare Green 
function (bare Cooperon); 
\\
{\it (iii)} is determined by factorizable product
of the ultraviolet divergent integrals. 

As we have already explained
such answer is hardly feasible for a one particle irreducible self
energy. In the second order of the
perturbation theory, there is an object which has a chance to
possess such a structure, see Fig.~1(d). However, it does not belong
to the {\em self-energy}.

{
\begin{figure}[ht]  
\epsfxsize= 7.5 cm  
\centerline{\epsfbox{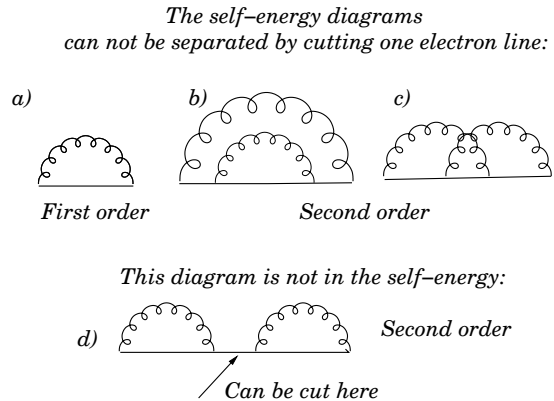}} 
\caption{First (a) and second order (b-d) diagrams.}
\vspace{0.2cm} 
\end{figure} 
}

{\it Conclusion} --- The equation for the self energy presented by GZ 
{\it (i)} has a well defined perturbative expansion in the interaction and
{\it (ii)} contradicts the conventional diagrammatic methods.
Therefore, the arguments of GZ about insufficiency of the perturbative
expansion lack the substance.  

Such a discussion could
be only justified, provided that GZ explicitly evaluate the
irreducible diagrams of topology of Fig.~1(b-c), 
and demonstrate that $1/\omega$
divergence appears in a double ultraviolet integral.
We do not require a full scale second
order calculation of all the diagrams with numerical coefficients,
 we would like to see  at least one  {\bf irreducible} graph which
is double ultraviolet and $1/\omega$ divergent at the same time.

\end{document}